\def\<{\langle} \def\>{\rangle}
\def\br{\bf\rm}  
\def\Tr{{\rm Tr}} \def\half{{\scriptstyle{1\over 2}}} 
   \def\M{{\br M}}  \def\d{{\rm d}}   \def\D{\Delta}
 \def\hs1{\hskip1mm} \def\h10{\hskip10mm}
\def\vs5{\vskip5mm} \def\page{\vfill\eject}

\def\ket#1{|#1\rangle} \def\bra#1{\langle#1|}
\def\expect#1{\langle#1\rangle}

\def\a{{\bf a}}  \def\Q{{\bf Q}}  \def\P{{\bf P}}  \def\Id{{\bf I}}
\def\H{{\bf H}}  \def\brho{{\bf\rho}}   \def\D{{\bf D}}  \def\L{{\bf L}}
\def\G{{\bf G}}

\documentstyle[12pt]{article}

\begin{document}

\title{Quantum state diffusion, localization and computation}

\author{R\"udiger Schack, Todd A Brun and Ian C Percival \\
  Department of Physics \\
  Queen Mary and Westfield College, University of London \\
  Mile End Road, London E1 4NS, England.}

\date{\today}
\maketitle

\begin{abstract}
Numerical simulation of individual open quantum systems has proven
advantages over density operator computations.  Quantum state
diffusion with a moving basis (MQSD) provides a practical numerical
simulation method  which takes full advantage of the localization of
quantum states into wave packets occupying small regions of classical
phase space.  Following and extending the original proposal of Percival,
Alber and Steimle, we show that  MQSD can provide a further gain over
ordinary QSD and other quantum trajectory methods of many orders of
magnitude in computational space and time. Because of these gains, it is
even possible to calculate an open quantum system trajectory when the
corresponding isolated system is intractable. MQSD is particularly
advantageous where classical or semiclassical dynamics provides an
adequate qualitative picture but is numerically inaccurate because of
significant quantum effects. The principles are illustrated by
computations for the quantum Duffing oscillator and for second harmonic
generation in quantum optics. Potential applications in
atomic and molecular dynamics, quantum circuits and quantum
computation are suggested. \end{abstract}

\vfill

J Phys A \hfill  QMW preprint Th--95--19

\page

\section{Introduction}

Most quantum systems are not even approximatedly isolated, but open, so
that they are signficantly affected by the environment.  This interaction is
important for atoms and molecules in gaseous or condensed matter
environments, which broaden spectral lines. It affects the motion of
molecules and  the rates of chemical reactions.  It is important for signalling
near the quantum limit, where the environment produces the noise
through which the signal must be detected. And it is important in
quantum optics, where it produces the dissipation that destroys
coherence.

Quantum state diffusion with a moving basis (MQSD) is a method of
representing and computing the evolution of individual open
quantum systems.  It has already been used by Percival et al.
\cite{Percival1995a} to analyse the motion  of a particle in a Penning
trap.  Here we provide a general theory of the method, and provide a
guide as to when it should be used in preference to other methods.
Second harmonic generation in optics and the quantum Duffing
oscillator are used as illustrations.

Because stochastic environmental fluctuations affect the evolution
of an individual open quantum system, it is represented  traditionally
by a density operator $\brho$, which
satisfies a linear master equation.  No attempt is made to represent
the evolution of individual pure states explicitly.  This approach is
adequate when the master equation has analytic solutions,  or when
the number of basis states $N$ of Hilbert space
required for numerical solution is not
too large.  But for large $N$ it often breaks down in practice long
before the corresponding numerical solution of the Schr\"odinger
equation, because the number of elements of a density matrix
increases as $N^2$.

Numerical simulation of individual open quantum systems, represented
by pure states which move along quantum trajectories, has proven
advantages over density operator computations.  Quantum state diffusion
(QSD) provides such a numerical simulation, in which each state diffuses
continuously in the state space and satisfies a nonlinear Langevin-It\^o
diffusion equation, determined uniquely by the master equation as shown in
\cite{Gisin1992c} and described in section \ref{secqsd}.  This
diffusion often produces a localization of quantum states into wave
packets that occupy small moving regions $R$ of classical phase space.
Such localization is a special characteristic of QSD that is usually
absent in other quantum state simulation methods
\cite{Garraway1994c}.

For a system of $f$ freedoms, a Planck cell of volume $(2\pi\hbar)^f$
corresponds to one quantum state.  The number of basis states used to
represent a system need not be much greater than the number of Planck
cells in the region $R$, {\it provided} that the basis follows the
motion of the wave packet in phase space.  Quantum state diffusion
with a moving basis (MQSD) provides a practical numerical simulation
method which takes full advantage of the localization, by referring
the quantum state to a moving origin $(q,p)$ in phase space.  This
origin lies at the phase space centroid of the quantum state,
determined by the current quantum expectations $\expect{\bf Q}$ and
$\expect{\bf P}$.

Numerical methods for solving the time-dependent Schr\"odinger equation
for an isolated system using moving wave packets have long been used
in chemical physics \cite{Heller1975,Lee1982,Coalson1982,Kucar1989}.   But
as the ordinary Schr\"odinger equation disperses wave packets instead of
localizing them in phase space, the applicability of these wave packet
approaches is very restricted.

For some systems, computation using MQSD is orders of magnitude more
economical in computer storage space and computation time than other
quantum state simulation methods, or the solution of master equations
for the density operator.  We give examples in which it is very difficult to
see how any other current method of numerical solution could be used.

Section \ref{secqsd} presents the basic QSD equations and their
derivation. The problems of the choice of boundary between system and
environment are described. There is a
brief comparison with other quantum trajectory methods, sometimes
called quantum jump or relative state methods
\cite{Carmichael1993b,Dalibard1992,Dum1992a,Gardiner1992}.
Localization is defined and discussed in section \ref{secloc}, with
reference to localization theorems and numerical examples. This is
followed by the definition of the moving basis, using excited coherent
states, and the derivation of the MQSD equations.

Section \ref{secbsp} applies MQSD to two challenging examples, the
Duffing oscillator and second harmonic generation.  These two examples
show how effective MQSD can be in bridging the gap between those
quantum problems where the number of basis states required in a fixed
basis is relatively small, and the quasiclassical limit where the
number of {\it fixed} basis states would be so large as to rule out
any practical use.  This section is completed by a crude analysis of
comparative computing times for the solution of an open system problem
using MQSD and the numerical solution of the Schr\"odinger equation
for a similar isolated system.

Section \ref{secconc} concludes with a comparison of methods for open
systems and some recommendations for their use, followed by the prospects
for using  MQSD in various applications.

\section{Quantum state diffusion (QSD)}  \label{secqsd}

Quantum state diffusion represents the evolution of a quantum
system through a correspondence between the solutions of the
master equation for the ensemble density operator $\brho$ and the
solutions of a Langevin-It\^o diffusion equation for the normalized
pure state vector $|\psi\>$ of an individual system of the ensemble
\cite{Gisin1992c}.

An analogy is helpful. A solution of the Langevin-It\^o equation for
the motion of an individual Brownian particle in position space
represents a single member of an ensemble whose distribution function
satisfies the corresponding Fokker-Planck equation.  Similarly, a
solution of the QSD equation for the diffusion of a pure quantum state
in state space represents a single member of an ensemble whose density
operator satisfies the corresponding master equation.  In each case
the effect of the environment can be represented either by the
stochastic evolution of an individual system or by the deterministic
evolution of the distribution.  For Brownian motion the evolution of
the individual system gives a more detailed picture of what happens to
an individual particle than the evolution of the distribution
function. Similarly, for open quantum systems, the evolution of the
individual pure states of QSD gives a more detailed picture of what
happens to an individual quantum system than the evolution of the
density operator.  This is particularly important for applications
like single particle traps, quantum noise in gravitational wave
detection, quantum circuits and quantum computers.

In QSD, {\it quantum expectations} $\<\dots\>$ for individual systems,
and {\it ensemble means} $\M$ are distinct.  The traditional quantum
expectation $\Tr \brho \G$ of an operator $\G$ for a mixed state of an
open system is equivalent in QSD to an ensemble mean over the quantum
expectations of the pure states $\ket{\psi}$.  That is
\begin{equation}
\Tr \brho \G = \M \<\G\> = \M \bra{\psi}\G\ket{\psi}.
\end{equation}

If the master equation has the standard Lindblad \cite{Lindblad1976} form
\begin{equation}
\dot\brho =
-{i\over\hbar}[\H,\brho] + \sum_m\left(\L_m\brho \L_m^{\dagger}
- \half \L_m^{\dagger}\L_m\brho   - \half\brho
\L_m^{\dagger}\L_m\right),
\label{eqmaster}
\end{equation}
then the corresponding QSD equation is a nonlinear
stochastic differential equation for the normalized state vector $|\psi\>$
of the ensemble, whose general differerential form is
\begin{eqnarray}
 |\d\psi\> = &
-{i\over\hbar} \H|\psi\>\d t + \sum_m\big(\<\L_m^{\dagger}\> \L_m
-\half \L_m^{\dagger} \L_m - \half\<\L_m^{\dagger}\>
\<\L_m\>\big) |\psi\>\d t \nonumber \\
  &+\sum_m \big(\L_m -\<\L_m\>\big) |\psi\>\d\xi_m,
\label{eqqsd}
\end{eqnarray}
where $\H$ is a Hamiltonian and $\L_m$ are
Lindblad  operators which represent the effect
of the environment on the system in a Markov approximation.

The first sum in (\ref{eqqsd}) represents the `drift' of the state
vector in the state space and the second sum the random fluctuations.
The $d\xi_m$ are independent complex differential random
variables, with normalized independent white noise in their
real and imaginary parts, leading to an
isotropic Brownian motion or Wiener process in the complex
$\xi_m$-plane.  These satisfy the conditions
\[
\M \d\xi_m = 0
\]
\begin{equation}
\M \d\xi_n \d\xi_m  = 0, \h10 \M \d\xi_n^* \d\xi_m     =
\delta_{nm}\d t,
\end{equation}
where $\M$ represents a mean over the ensemble.  The complex Wiener
process is normalized to $\d t$, so the independent real and imaginary Wiener
processes are each normalized to $\d t/2$.  This is the same
normalization as  in \cite{Gisin1993d,Gisin1993b,Percival1994b},
but is different from that of \cite{Gisin1992c}.  The distribution (4) is
invariant
under unitary transformations in the linear space of the $\d\xi_m$.

$\<\L_m\> = \<\psi|\L_m|\psi\>$ is the
quantum expectation of $\L_m$ for state  $|\psi\>$. The density operator
is given by the  mean over the projectors onto the quantum states of the
ensemble:
\begin{equation}
\brho = {\M}|\psi\>\<\psi|.
\label{eqrho}
\end{equation}
 It can be verified that if the pure states of the ensemble satisfy
the QSD equation (\ref{eqqsd}), then the density operator
(\ref{eqrho}) satisfies the master equation (\ref{eqmaster}).  There
are many diffusion equations for pure states $\ket{\psi}$ which give
the same master equation for the density operator.  The uniqueness of
the QSD equations follows from a principle of unitary invariance in
operator space.

This is most clearly illustrated by a QSD equation with a single
Lindblad operator $\L$. In that case unitary transformation in
operator space is multiplication by a scalar phase factor $u$ of
modulus unity. Then it is obvious by inspection that if $\L$ is
replaced by $u\L$,  the master equation, and hence the solution of
the master equation, is unchanged. In the corresponding QSD
equation, the replacement of $\L$ by $u\L$ is the same as the
replacement of $\d\xi$ by $u\d\xi$, but since the distribution of the
elementary differential fluctuations $\d\xi$ is invariant for
multiplication by a phase factor, the QSD equations are also
unchanged.  This is not true if the fluctuations are real or otherwise
do not satisfy unitary invariance in fluctuation space, as in
\cite{Ghirardi1986,Gisin1990}. The QSD equations
are the only diffusion equations that respect this unitary invariance
property in the one-dimensional operator space,  and they are in this
respect unique up to a physically irrelevant external time-dependent
phase factor for the state vector.

This result generalizes to an arbitrary number of Lindblad operators
if $u$ is taken to be a general unitary transformation in the linear
space of the Lindblad operators.  The general result was first given
for QSD equations in \cite{Gisin1992c},  following the statement of
an invariance principle by Di\'osi \cite{Diosi1988a}, and the
derivation for a Fokker-Planck equation in state space in
\cite{Percival1989}.

In either the traditional or QSD formulation for open systems,  there
is always a problem as to where to put the boundary between system
and environment.  If the system is made too small, then important
effects are neglected, and errors are made.  If it is made too big by
including too much of the environment, then the dimension $N$ of
the basis state space becomes too large for the equations to be
solved.  In practice there is a compromise.  Often, in addition, the
problem is simplified by approximating the effects of complicated
environments  by simple operators.

Because of the localization property discussed in the next section,
QSD has the property that it has no need of a separate
measurement hypothesis, as shown in detail with examples in
\cite{Gisin1992c,Gisin1993d,Gisin1993b}.  Measuring apparatus
is just one type of environment, and its effects can be represented
by simple operators.
Alternatively, measurement can be represented in detail by pushing
the boundary between system and environment out so far that the
system includes the measuring apparatus.  This is useful in
establishing the representation of measurement by operators, but
results in a system too complicated to solve directly.
Localization and the representation of measuring apparatus are needed for
comparison with the relative state or quantum jump methods.

QSD theory followed from research that was motivated by the desire
to find an explicit physical representation of the measurement
process. Following pioneering work of Bohm and Bub \cite{Bohm1966} and
Pearle \cite{Pearle1976,Pearle1979}, Gisin \cite{Gisin1984a} introduced a
simple example of quantum state diffusion with real fluctuations
that was generalized by Di\'osi \cite{Diosi1988a} and Gisin \cite{Gisin1989a}.
The complex It\^o form of QSD was introduced in
\cite{Gisin1992c}.  The detailed QSD theory and its applications are
described in \cite{Gisin1992c,Gisin1993d,Gisin1993b,Percival1994b}.
Diffusion in the space of quantum states also
appears in connection with the theory of continuous quantum
measurements as shown,  for example, in the many references in
\cite{Barchielli1991,Carmichael1993b}.

Gisin and Percival \cite{Gisin1993c} used QSD to describe a quantum
jump experiment. Goetsch and Graham \cite{Goetsch1993} used it to
describe some nonlinear optical processes.  Garraway and Knight
\cite{Garraway1994a} compared QSD and quantum jump simuations for
two-photon processes, and in \cite{Garraway1994b} they compared the
phase space picture of QSD and quantum jumps, showing that the former
gives localization and the latter does not.  Spiller and his coworkers
applied QSD to thermal equilibrium \cite{Spiller1993a}, studied chaos
in a simple open quantum system \cite{Spiller1994a} and investigated
open angular systems, such as quantum capacitors and
rotors \cite{Spiller1995a}.  Gisin \cite{Gisin1993e} investigated the
Heisenberg picture for QSD.  Further examples are given in
\cite{Gisin1992c,Gisin1993d,Gisin1993b,Salama1993}.

\section{Localization and the moving basis}  \label{secloc}

For a quantum system in a pure state, localization refers to dynamical
variables and operators like position, momentum, energy and angular
momentum which have classical equivalents.  It is helpful to picture
the localization in classical phase space.  There are many phase space
representations of quantum systems, such as the Wigner distribution,
but these are not necessary to obtain a classical
picture of the phase space localization of a quantum system.  For
that, it is only necessary to consider the expectations and variances
of quantum dynamical variables, and then to picture these expectations
and variances {\it as if} they represented the expectations and
variances of classical quantities.

The general theory of localization is treated in
\cite{Gisin1992c} and in \cite{Percival1994b}.  The
Schr\"odinger evolution of a system usually produces {\it de}localization
or dispersion. The interaction of the system with its environment, by
contrast, produces localization.

It is convenient to consider these competing processes separately
before considering them together.  The dispersion is well-known,
since it occurs in isolated systems.  The opposite extreme is an open
system which interacts with environment so strongly that the
Schr\"odinger evolution can be neglected.  This is a {\it wide open}
system, and the theory of localization for these systems has
been treated in detail in \cite{Percival1994b}.

For simple systems, earlier
papers \cite{Gisin1992c,Gisin1993d,Gisin1993b} had
built up a picture in which interaction with the environment produced
localization, sometimes to an eigenstate of an operator
corresponding to a surface in phase space, but more commonly to a
state which is localized to a wave packet in phase space whose
Heisenberg indeterminacy products are of the order of
 $\hbar$. In \cite{Percival1994b} that picture was confirmed and
extended.  A general theory was presented, lower bounds were put
on rates of self-localization, and bounds were put on asymptotic
states.

For a pure state $\ket{\psi}$, the expectation of a selfadjoint
operator $\G$ is denoted
\begin{equation}
\expect{\G} = \<\psi|\G|\psi\>
\end{equation}
and the variance of the operator $\G$ is
\begin{equation}
\sigma^2(\G) = \expect{\G^2} - \expect{\G}^2.
\end{equation}
The localization $\Lambda$ is defined as the inverse of the mean of
the variance for the ensemble:
\begin{equation}
\Lambda = \big(\M\sigma^2(\G)\big)^{-1}.
\end{equation}
The simplest theorem is for a wide open system in
which there is only one self-adjoint Lindblad operator $\L$ in the state
diffusion equation.  It is shown that the rate of localization of $\L$
is at least as fast as 2 (\cite{Percival1994b}, section \ref{secloc}):
\begin{equation}
\d\Lambda/\d t \ge 2.
\end{equation}
Because Lindblad operators have dimension ${\rm time}^{-\half}$,
the rate is dimensionless.

This localization is towards a {\it surface} in phase space defined by
the dynamical variable $L$, and is characteristic of interaction with
apparatus that measures $L$.  More general and more interesting is
the case where there are at least $2f$ operators, where $f$ is the
number of freedoms, and the surfaces for the corresponding
dynamical variables define points in phase space. In that case at
least $f$ pairs of operators will not commute.  The simplest
example is a one freedom system with two conjugate variables $X,Y$
whose operators satisfy
\begin{equation}
[{\bf X},{\bf Y}] = i\hbar.
\end{equation}
In this case it is shown that the
state  localizes and asymptotically approaches a wave packet with
minimum Heisenberg indeterminacy product. It is also shown that
operators that are not selfadjoint, such as annihilation and creation
operators, can localize to wave packets. In the more realistic
examples that we study numerically in the
following section, the localization has to
compete with the dispersion due to the Schr\"odinger evolution,
 the wave packet is more dispersed, and its dispersion often varies
considerably as a function of time.

For general open systems there is a physical competition between
the dispersion due to the Hamiltonian and the localization due to the
state diffusion.  There is no general theory for this, but there are
many numerical examples of localization, such
as \cite{Gisin1992c} for the forced damped oscillator and
\cite{Gisin1993b} for localization in one well of a double well, and
phase space localization by position localization and Hamiltonian
coupling of position and momentum. Halliwell and
Zoupas \cite{Halliwell1995a} have provided a general theory for the
evolution of Gaussian wave packets for an important model,
generalizing a result of Di\'osi \cite{Diosi1988c}.

{}From the theorems and the numerical examples it would appear that
the  localization of wave packets to  relatively small regions of
phase space is the norm, so that the  limiting behaviour on a classical
scale is the direct representation of classical states as points in
phase space, rather than the more abstract surfaces defined by
action functions that satisfy the Hamilton-Jacobi equation.

For our purposes the most important consequence of the
localization of quantum trajectories around phase space trajectories
is, that by continually changing the basis, it is often possible to
reduce the number of basis states needed to represent the wave
packet by many orders of magnitude. If a wave packet is localized
about a point $(q,p)$ in phase space far from the origin, it requires
a great many of the usual number states $\ket{n}$ to represent it.
But  fewer {\it excited coherent} basis states $\ket{q,p,n} =
\D(q,p)\ket{n}$, are needed, with corresponding savings in
computer storage space and computation time.  These states
are defined using the coherent state displacement operator,
\begin{equation}
\D(q,p) = \exp{i\over\hbar} \biggl( p\Q - q\P \biggr).
\end{equation}
The separation of the representation into a classical part $(q,p)$
and a quantum part $\ket{q,p,n}$ is called the {\it moving basis}, or,
as in \cite{Percival1995a}, the {\it mixed} representation.

In this basis, the usual creation and annihilation operators are
modified:
\begin{equation}
 \a = \a(q,p) + (q + i p)\Id/\sqrt{2},\h10
\a^\dagger = \a^\dagger(q,p) + (q - i p)\Id/\sqrt{2},
\end{equation}
where $\a(q,p)$ is the local annihilation operator
(with $(q,p)$ as the origin) and $\Id$ is the identity operator. The
effect of the local operators on the excited coherent states
is given by:
 \begin{equation}
\a(q,p)\ket{q,p,n} = \sqrt{n}\hs1\ket{q,p,n-1},\h10
\a^\dagger(q,p)\ket{q,p,n} = \sqrt{n+1}\hs1\ket{q,p,n+1}.
\end{equation}
Similarly,
\begin{equation}
\Q = \Q(q,p) + q\Id, \h10 \P = \P(q,p) + p\Id,
\end{equation}
where
\begin{equation}
\Q(q,p) = (\a(q,p) + \a^\dagger(q,p) )/\sqrt{2}, \h10
  \P(q,p) = - i (\a(q,p) - \a^\dagger(q,p) )/\sqrt{2}.
\end{equation}

Care must be taken to avoid ambiguity in the external phase factors.
Normally, different displacement operators do not commute; nor is the
product of two displacement operators a standard displacement operator.  In
both cases there is an additional phase factor:
\begin{equation}
\D(q^\prime,p^\prime)\ket{q,p,n} =
  \ket{q+q^\prime,p+p^\prime,n} {\rm e}^{i(qp^\prime - pq^\prime)/2}.
\end{equation}
In order to retain an unambiguous relation to the
standard fixed basis states, this additional phase factor must be removed.
Fortunately this is simple, as it the same for all the moving basis states.

The numerical algorithm follows directly.  As the integration proceeds,
the phase point expectation
\begin{equation}
\big(\delta q, \delta p\big) =
  \big(\expect{\Q(q,p)},\expect{\P(q,p)}\big)
\end{equation}
drifts away from zero.  The basis is then shifted,
\begin{equation}
\ket\psi \rightarrow \D(-\delta q,-\delta p) \ket\psi,
\end{equation}
and the classical phase
point $(q,p)$ adjusted:
\begin{equation}
(q,p) \rightarrow \big( q + \delta q, p + \delta p\big).
\end{equation}
The time between basis adjustments can be optimized in various
ways, depending on the problem, the degree of localization, and the
difficulty of carrying out the basis change.

In principle there is another way to make the basis change.
If the phase space point
\begin{equation}
(q,p) =(\expect{\Q},\expect{\P})
\end{equation}
and state $\ket\psi$ {\it relative to} (q,p) {\it as origin} are used
to define the state, then we can include the change of basis in the
evolution equation for $\ket\psi$. This leads to a set of
simultaneous differential equations in $q$, $p$, and $\ket\psi$; in
integrating these equations, both the QSD evolution and the basis
shift are done automatically.  This {\it simultaneous moving basis}
method is attractive for a number of reasons; the states always
remain in an optimally `localized' basis, and therefore can take the
best advantage of the QSD localization effects to minimize the
number of necessary basis states.   There is no need
for an additional step for the basis change.  In practice, the
simultaneous moving basis method results in a considerable increase in
programming complexity, and may have numerical
stability problems.  As yet the method has not
proved itself sufficiently, so we have not used it.

Implementing the moving basis algorithm on a computer is
straightforward. Each set of normalized flucutations $\d\xi_m$
determines a quantum trajectory through Eq.~(\ref{eqqsd}), which can be
simulated using discrete time steps $\delta t$,  using a
Runge-Kutta algorithm for the deterministic part, and an Euler
method for the stochastic part (but see also \cite{Steinbach1994}).
Suppose that at
time $t=t_0$ the state $\ket{\psi(t_0)}$ is represented in the
basis $\ket{q_0,p_0,n}$, centered at
\begin{equation}
(q_0,p_0) = \big(\expect{\psi(t_0)|\Q|\psi(t_0)},
\expect{\psi(t_0)|\P|\psi(t_0)}\big).
\end{equation}
Then after one discrete time step, the expectations in this basis shift to
\begin{equation}
(q'_0,p'_0) = \big(\expect{\psi(t_0+\delta t)|\Q|\psi(t_0+\delta
t)},
\expect{\psi(t_0+\delta t)|\P|\psi(t_0+\delta t)}\big)\ne
(q_0,p_0).
\end{equation}

 The computational advantage of a small number of basis states
is then retained by changing the representation
to the shifted basis $\ket{q_1,p_1,n}$ centered at $q_1$ and $p_1$.
This shift in the origin of the basis represents the elementary single step
of the moving basis of MQSD.

The components of $\ket{\psi(t_0+\delta t)}$ can be computed using the
expressions given above. The computing time needed for the
basis shift is of the same order of magnitude as for computing a
single discrete timestep of Eq.~(\ref{eqqsd}).  Shifting the basis
once every discrete timestep could therefore double the computing
time, depending on the complexity of the Hamiltonian and the number of
degrees of freedom. On the other hand, the reduced number of basis
vectors needed to represent states in the moving basis can lead to
savings far bigger than a factor of 2.

In one example of second harmonic generation described in the
following section, two modes of the electromagnetic field
interact. Using the moving basis reduces the number of basis vectors
needed by a factor of 100 in each mode. The total number of basis
vectors needed is thus reduced by a factor of 10000, leading to
reduction in computing time by a factor of $10000/2=5000$. Futhermore,
the fixed basis would exceed the memory capacity of most existing
computers.

The QSD equation~(\ref{eqqsd}) can contain both localizing and
delocalizing terms.  Nonlinear terms in the Hamiltonian tend to spread
the wave function in phase space, whereas the Lindblad terms
localize. Accordingly, the width of the wave packets varies along a
typical trajectory. We use this to reduce the computing time even
further by dynamically adjusting the number of basis vectors. Our
criterion for this adjustment depends on parameters $\epsilon\ll1$,
the {\it cutoff probability}, and $N_{\rm pad}$, the {\it pad size},
which represents the number of boundary basis states that are checked
for significant probability. We require the total probability of the
top $N_{\rm pad}$ states to be no greater than $\epsilon$, increasing
and decreasing the number of states actually used accordingly, as the
integration procedes along the quantum trajectory.

\section{Examples:  The Duffing oscillator and second harmonic generation}
\label{secbsp}

The quantum mechanics of systems whose classical limit exhibits
dissipative chaos is an interesting problem.  Dissipation is
relatively difficult to treat in quantum mechanics.  The best of the
commonly used techniques is the solution of the master equation, but
as we have pointed out, solving the master equation numerically can be
an extremely difficult problem.  Perhaps because of this, quantum
dissipative chaos has been neglected by comparison with quantum
hamiltonian chaos. Clearly the QSD method for open systems is
eminently suitable for such dissipative systems. Recently Spiller and
Ralph \cite{Spiller1994a} applied QSD to a dissipative chaotic system.

The Duffing oscillator is particularly appropriate for applying MQSD,
as the classical system has been widely studied \cite{Guckenheimer1983}.
This system has also been treated quantum mechanically in the decoherent
histories formalism \cite{Brun1993b,Brun1994},
which has been shown to have connections
to QSD \cite{Diosi1995}.
This oscillator consists of a particle moving in one dimension in the
two-welled potential
\begin{equation}
V(x) = {{x^4}\over4} - {{x^2}\over2}.
\end{equation}
Dissipative chaos can be produced by adding both dissipation
(of the form $-2\Gamma{\dot x}$) and a
periodic driving force (of the form $g\cos(t)$).

By using either the master equation or QSD, it is relatively simple to
calculate the evolution of this system far from the classical limit, for
example when $\hbar = 1$.  But it is more important to
study the classical limit of quantum chaos, to see how the relatively
well understood properties of classical chaos appear in quantum
systems.  This limit is conveniently represented by decreasing
$\hbar$, which increases the number of fixed basis states required.
With $\hbar = 1$, about 10 states are needed to simulate
the system with accuracy.  On approaching the classical limit, the
number of states quickly becomes impractical.  In a semiclassical
regime with $\hbar \approx 10^{-4}$, more than 10,000 states are
needed; the density operator method would require storage and
computation with a prohibitive $10^8$ real numbers. This, like most
nonlinear problems, is exacerbated by the fact that the
potential needs much more computation than, for example, the
simple harmonic oscillator.

Using MQSD, however, this problem becomes tractable.  In a chaotic
system, the delocalizing forces are particularly strong in the
neighbourhood of hyperbolic fixed points, where the dynamics spreads
the wave packet in phase space.  However, the localizing effects of
the environment always predominate.  MQSD picks out a good local
time-dependent basis, and the method rarely requires more than twenty
states, usually about ten, just as in the small scale quantum limit.
This is easy to see by examining figure \ref{figduff}.

Our second example is frequency doubling or second harmonic
generation, which is a standard process in quantum optics.  The system
consists of two optical modes of frequency $\omega_1$ and
$\omega_2\simeq2\omega_1$ which interact in a cavity driven by a
coherent external field with frequency $\omega_f\simeq\omega_1$ and
amplitude $f$. The cavity modes are slightly damped and detuned, with
detuning parameters $\delta_1=\omega_1-\omega_f$ and
$\delta_2=\omega_2-2\omega_f$.

The Hamiltonian in the interaction picture is
\begin{equation}
\H = \hbar \delta_1 \a_1^\dagger \a_1  +  \hbar \delta_2 \a_2^\dagger \a_2
    + i\hbar f (\a_1^\dagger-\a_1)
    + i\hbar{\chi\over2} (\a_1^{\dagger2} \a_2 - \a_1^2 \a_2^\dagger) \;,
\end{equation}
where $\a_1$ and $\a_2$ are the annihilation operators of the two
cavity modes, and $\chi$ describes the strength of the nonlinear
interaction between them.  Damping of the two cavity modes is
described by the Lindblad operators $\L_1=\sqrt{2\kappa_1}\a_1$ and
$\L_2=\sqrt{2\kappa_2}\a_2$. The factors of $\sqrt{2}$ are a
consequence of normalization conventions used in the master equation
Eq.~(\ref{eqmaster}) which differ from those commonly used in quantum
optics.  The master equation for this problem first appeared in
\cite{Drummond1980b,Drummond1981a}.

A direct numerical solution of the master equation for this problem is
difficult because, in the Fock basis, the dimension of the effective
Hilbert space is equal to the product of the photon number cutoffs in
both modes and easily becomes very large. The problem becomes
intractable even for moderate numbers of photons in each mode. Earlier
treatments of this problem include
\cite{Doerfle1986,Savage1988,Schack1991c,Goetsch1993,Zheng1995}.

The QSD method was employed for studying second harmonic generation by
Gisin and Percival \cite{Gisin1992c} and Goetsch and Graham
\cite{Goetsch1993}. Both used a fixed Fock-state basis, and rather
limited photon numbers.  Gisin and Percival \cite{Gisin1992c} improved
the method slightly by using a lower as well as an upper cutoff. More
recently, Zheng and Savage \cite{Zheng1995} applied the quantum jump
method to the chaotic regime \cite{Savage1983} of second-harmonic
generation for large photon numbers. Using hundreds of hours
on a 32-processor supercomputer, they succeed in simulating cases
where the expectation of the photon number is of the order of 200 in
each mode. Their Fock basis has 512 basis states in each mode, i.e., a
total of 250000 basis states, a {\it tour-de-force\/} near the limits
of their method.

To illustrate the power of MQSD, we have computed trajectories for
second harmonic generation with parameters similar to the ones used by
Zheng and Savage \cite{Zheng1995}, but leading to photon numbers on
average six times as large. For this problem, Zheng and Savage would
have needed approximately $36\times250000\simeq10^7$ basis
states. Using the moving basis with an adaptive basis size, we
computed a single trajectory in a few hours on a PC. The results shown
in Fig.~\ref{figshg} were obtained using a cutoff probability
$\epsilon=10^{-3}$. The total number of basis states needed in the
simulation varied between approximately 50 and 1000. Changing the
cutoff probability to $\epsilon=10^{-2}$ lowered the number of basis
states needed to between approximately 40 and 200, clearly a
strong dependence on $\epsilon$.  Surprisingly, the choice of cutoff
probability affected the simulated expectation values only very
slightly.

It is interesting to compare estimates of computation space and time
of a single MQSD run for an open sytems and the numerical solution
of the Schr\"odinger equation for the same system isolated from its
environment.  The former has diffusion and drift terms that increase
the computation time for a single step of integration, but the
localization  reduces significantly the size of the basis needed for
computation. The latter has fewer terms in the equation, but the
Schr\"odinger dispersion increases the number of states needed for
the representation.  Clearly, for systems with more than one degree
of freedom the advantage of MQSD is even greater, the savings in
number of states going like the power of the number of freedoms.
We can make this semi-quantitative by the following analysis:

Consider computing times. For each computation, let $N_j$ be the
arithmetic mean of the number of basis states required for the $j$th
freedom.   Let $\bar N$ be the geometric mean of all the $N_j$,
which is is an appropriate measure for the number of basis states for
one freedom. Let $f$ be the number of freedoms. Then an estimate
of the computing time $T_{\rm comp}$ is given by $K {\bar N}^f$,
where the constant $K$ is much larger for MQSD but $\bar N$ is
much larger for Schr\"odinger.  The ratio of computing times is then
\begin{equation}
{T_{\rm comp}({\rm Sch})\over T_{\rm comp}({\rm MQSD})}
\approx {K({\rm Sch})\over K({\rm MQSD})}
\Big[{\bar N({\rm Sch})\over \bar N ({\rm MQSD})}\Big]^f.
\label{timing}
\end{equation}

The values of these numbers are highly problem-dependent.  Clearly,
for a problem with a complicated Hamiltonian and one simple
Lindblad operator (as in the Duffing oscillator), the difference in
time between the QSD and Schr\"odinger calculations will be much
less than it would be for a simple Hamiltonian with several Lindblad
operators (as in second harmonic generation).

One can estimate the values of these numbers for the problems described.
For the Duffing Oscillator, $K({\rm Sch})/K({\rm MQSD}) \approx 2/3$,
$\bar N({\rm Sch})/\bar N({\rm MQSD}) \approx 1500$, and
$f = 1$.  From (\ref{timing}) we expect that the ratio of computing
time between MQSD and the Schr\"odinger equation is roughly
$10^{-3}$. In the case of second harmonic generation, we
have $K({\rm Sch})/K({\rm MQSD}) \approx 1/2$,
$\bar N({\rm Sch})/\bar N({\rm MQSD}) \approx 100$ and $f=2$, yielding
a ratio of computing times of approximately $2 \times 10^{-4}$.
Clearly, the advantage for multi-freedom problems is very
great.

\section{Conclusions and prospects}   \label{secconc}

For the simplest open system problems, where analytic solutions are
available, or the number $N$ of basis states that are needed for the
solution of the master equation is small, a density operator method is to be
preferred.  This method may be the only one avaibable when very high
accuracy is needed, because of the inevitable statistical errors in Monte
Carlo simulations.

Otherwise a  method that simulates individual states is to be preferred.

Where the main interaction of the open system is with a measuring
apparatus, which records the quantum state by counting, as with a
photon counter, then the relative state or quantum jump method is
the natural one and the best to use for numerical computations.  The
application of QSD produces rapid changes in state that look like
quantum jumps \cite{Gisin1992c, Gisin1993c}, but this  is
numerically inefficient.

When measurement is not the main interaction with an environment,
then QSD is often more efficient.   Examples are the types of thermal
interaction that are often represented by heat baths,
interactions with a radiation field, and also continuous interactions
with  condensed matter or gaseous environments that are not to be
considered as part of the system, as in line broadening,  molecular
dynamics,  noise in quantum circuits, etc.  In these cases it is often
simplest to use QSD with a fixed basis, and this is likely to be more
efficient numerically than jumps.

Where the interaction with the environment is sufficiently strong to
produce strong localization during a significant part of the
evolution, then MQSD is to be preferred.  This is a very common
situation, as the greater the environmental interaction, the greater
the localization becomes.  MQSD is feasible where there are several
degrees of freedom, and there would be no hope of using other
simulation methods. In some limiting cases semiclassical methods can
be used instead. MQSD is likely to be valuable in the those situations
that lie between those that are sufficiently simple to be simulated by
other numerical means, and those for which a semiclassical method
can be used.

Monte Carlo trajectory methods for open systems have been developed
largely in connection with quantum optics, but they have been extended
using QSD and MQSD to particles in traps \cite{Percival1995a},
while Spiller and his collaborators \cite{Spiller1995a} have
carried out some interesting QSD studies of quantum circuit elements
with a view to applications to Josephson junctions.

However, the potential range of useful applications is far greater.  In
particular there has been  no application to quantum effects in the
dynamics of molecules which interact strongly with gases or with
liquids. For isolated molecules, there are often transitions between
many potential energy surfaces, and no clear way of localizing the
molecule on one surface out of many.  But in QSD, the interaction with
the enviroment localizes the molecule into a small region around a
phase point on one many possible surfaces, so that MQSD for a molecule
interacting with its environment would have considerable advantages
over the numerical solution of the Schr\"odinger equation for an
isolated molecule.

It is also clear that the noise in quantum circuits is an
environmental effect that could well be simulated by MQSD.  This
applies even more strongly to quantum computation, where the
potentially rapid decoherence produced by thermal interaction could be
a crucial limitation.  In these cases a lot could be learned from
individual MQSD runs, without the need to make the large number of
simulations required to get good statistics.  It is not even clear
that the density operator is of much relevance to such applications,
for which the individual system is all-important.

\section*{Acknowledgments}

We thank the EPSRC in the UK and also the
University of Geneva for essential financial support, and G Alber, B
Garraway, N Gisin, J Halliwell, P Knight, T Steimle,  W Strunz and P
Zoller for valuable communications.

\newpage

\begin{figure}
\caption{Poincar\'e surface for
expectation values of $x$ versus $p$ for the forced,
damped Duffing oscillator, plotted at times of constant phase
$t_n = 2\pi n$.  This is in the chaotic regime, with constants
$g = 0.3$ and $\Gamma = 0.125$, and scaled up by a factor of 100
in $x$ and $p$, effectively reducing $\hbar$ by a factor of $10^4$.
The system is initially in the ground state.  Note that it would
take $\sim 50,000$ states to represent this system with a non-moving
basis.  The cutoff probability is $\epsilon = 5\times10^{-3}$.}
\label{figduff}
\end{figure}

\begin{figure}
\caption{Photon number versus dimensionless scaled time $\tau=\kappa_1t$ for a
single second harmonic generation trajectory. Solid line:
fundamental mode $\expect{\a_1^\dagger \a_1}$; broken line: second
harmonic mode $\expect{\a_2^\dagger\a_2}$. The parameters are
$\kappa_2/\kappa_1 = 0.25$, $\delta_1/\kappa_1=\delta_2/\kappa_1 =
-1$, $\chi/\kappa_1 = 0.5$, and $f/\kappa_1 = 62$. At time $t=0$, the
system is in the vacuum state. These parameters lie in the
chaotic regime of the corresponding classical system.
The cutoff probability is $\epsilon=10^{-3}$.}
\label{figshg}
\end{figure}


\begin{thebibliography}{10}

\bibitem{Barchielli1991}
{\sc Barchielli, A., and Belavkin, V.~P.}
\newblock {\em J. Phys.\ A 24} (1991), 1495.

\bibitem{Bohm1966}
{\sc Bohm, D., and Bub, J.}
\newblock {\em Rev.\ Mod.\ Phys.\ 38\/} (1966), 453.

\bibitem{Brun1993b}
{\sc Brun, T.~A.}
\newblock {\em Caltech Preprint\ CALT-68-1882\/} (1993).

\bibitem{Brun1994}
{\sc Brun, T.~A.}
\newblock {\em Applications of the Decoherence Formalism}.
\newblock Caltech Ph.D. Thesis, UMI Dissertation Services, 1994.

\bibitem{Carmichael1993b}
{\sc Carmichael, H.~J.}
\newblock {\em An Open Systems Approach to Quantum Optics}.
\newblock Springer, Berlin, 1993.

\bibitem{Coalson1982}
{\sc Coalson, R.~D., and Karplus, M.}
\newblock {\em Chem.\ Phys.\ Lett.\ 90\/} (1982), 301.

\bibitem{Dalibard1992}
{\sc Dalibard, J., Castin, Y., and M{\o}lmer, K.}
\newblock {\em Phys.\ Rev.\ Lett.\ 68\/} (1992), 580.

\bibitem{Diosi1988c}
{\sc Di\'osi, L.}
\newblock {\em Phys.\ Lett.\ A 132} (1988), 233.

\bibitem{Diosi1988a}
{\sc Di\'osi, L.}
\newblock {\em J. Phys.\ A 21\/} (1988), 2885.

\bibitem{Diosi1995}
{\sc Di\'osi, L., Gisin, N., Halliwell, J., and Percival, I.~C.}
\newblock {\em Phys. Rev. Lett.\ 21\/} (1995), 203.

\bibitem{Doerfle1986}
{\sc D\"orfle, M., and Schenzle, A.}
\newblock {\em Z. Phys.\ B 65\/} (1986), 113.

\bibitem{Drummond1980b}
{\sc Drummond, P.~D., McNeil, K.~J., and Walls, D.~F.}
\newblock {\em Optica Acta 27\/} (1980), 321.

\bibitem{Drummond1981a}
{\sc Drummond, P.~D., McNeil, K.~J., and Walls, D.~F.}
\newblock {\em Optica Acta 28\/} (1981), 211.

\bibitem{Dum1992a}
{\sc Dum, R., Zoller, P., and Ritsch, H.}
\newblock {\em Phys.\ Rev.\ A 45} (1992), 4879.

\bibitem{Gardiner1992}
{\sc Gardiner, C.~W., Parkins, A.~S., and Zoller, P.}
\newblock {\em Phys.\ Rev.\ A 46} (1992), 4363.

\bibitem{Garraway1994c}
{\sc Garraway, B.~M., and Knight, P.~L.}
\newblock {\em Phys.\ Rev.\ A 49} (1994), 1266.

\bibitem{Garraway1994a}
{\sc Garraway, B.~M., and Knight, P.~L.}
\newblock {\em Phys.\ Rev.\ A 50} (1994), 2548.

\bibitem{Garraway1994b}
{\sc Garraway, B.~M., Knight, P.~L., and Steinbach, J.}
\newblock {\em Appl.\ Phys.\ B 60} (1995), 63.

\bibitem{Ghirardi1986}
{\sc Ghirardi, G.~C., Rimini, A., and Weber, T.}
\newblock {\em Phys.\ Rev.\ D 34} (1986), 470.

\bibitem{Gisin1984a}
{\sc Gisin, N.}
\newblock {\em Phys.\ Rev.\ Lett.\ 52\/} (1984), 1657.

\bibitem{Gisin1989a}
{\sc Gisin, N.}
\newblock {\em Helvetica Physica Acta 62\/} (1989), 363.

\bibitem{Gisin1990}
{\sc Gisin, N.}
\newblock {\em Physica A 63} (1990), 929.

\bibitem{Gisin1993e}
{\sc Gisin, N.}
\newblock {\em J. Mod.\ Opt.\ 40} (1993), 2313.

\bibitem{Gisin1993c}
{\sc Gisin, N., Knight, P.~L., Percival, I.~C., Thompson, R.~C., and Wilson,
  D.~C.}
\newblock {\em J. Mod.\ Opt.\ 40} (1993), 1663.

\bibitem{Gisin1992c}
{\sc Gisin, N., and Percival, I.~C.}
\newblock {\em J. Phys.\ A 25} (1992), 5677.

\bibitem{Gisin1993d}
{\sc Gisin, N., and Percival, I.~C.}
\newblock {\em J. Phys.\ A 26} (1993), 2233.

\bibitem{Gisin1993b}
{\sc Gisin, N., and Percival, I.~C.}
\newblock {\em J. Phys.\ A 26} (1993), 2245.

\bibitem{Goetsch1993}
{\sc Goetsch, P., and Graham, R.}
\newblock {\em Ann.\ Physik 2} (1993), 706.

\bibitem{Guckenheimer1983}
{\sc Guckenheimer, J., and Holmes, P.}
\newblock {\em Nonlinear Oscillations, Dynamical Systems,and Bifurcations
  of Vector Fields}.
\newblock Springer-Verlag, New York, 1983.

\bibitem{Halliwell1995a}
{\sc Halliwell, J.~J., and Zoupas, A.}
\newblock Quantum state diffusion, density matrix diagonalization and
  decoherent histories: A model, 1995, to be published.

\bibitem{Heller1975}
{\sc Heller, E.~J.}
\newblock {\em J. Chem.\ Phys.\ 62\/} (1975), 1544.

\bibitem{Kucar1989}
{\sc Kucar, J., and Meyer, H.-D.}
\newblock {\em J. Chem.\ Phys.\ 90\/} (1989), 5566.

\bibitem{Lee1982}
{\sc Lee, S.~Y., and Heller, E.~J.}
\newblock {\em J. Chem.\ Phys.\ 76\/} (1982), 3035.

\bibitem{Lindblad1976}
{\sc Lindblad, G.}
\newblock {\em Commun.\ Math.\ Phys.\ 48\/} (1976), 119.

\bibitem{Pearle1976}
{\sc Pearle, P.}
\newblock {\em Phys.\ Rev.\ D 13\/} (1976), 857.

\bibitem{Pearle1979}
{\sc Pearle, P.}
\newblock {\em Int.\ J. Theor.\ Phys.\ 18\/} (1979), 489.

\bibitem{Percival1989}
{\sc Percival, I.~C.}
\newblock Diffusion of quantum states 2, 1989.
\newblock Preprint QMC DYN 89-4, School of Mathematics, Queen Mary College
  London.

\bibitem{Percival1994b}
{\sc Percival, I.~C.}
\newblock {\em J. Phys.\ A 27} (1994), 1003.

\bibitem{Percival1995a}
{\sc Percival, I.~C., Alber, G., and Steimle, T.}
\newblock submitted to J. Phys.\ A.

\bibitem{Salama1993}
{\sc Salama, Y., and Gisin, N.}
\newblock {\em Phys.\ Lett.\ A 181} (1993), 269.

\bibitem{Savage1988}
{\sc Savage, C.~M.}
\newblock {\em Phys.\ Rev.\ A 37\/} (1988), 158.

\bibitem{Savage1983}
{\sc Savage, C.~M., and Walls, D.~F.}
\newblock {\em Optica Acta 30} (1983), 557.

\bibitem{Schack1991c}
{\sc Schack, R., and Schenzle, A.}
\newblock {\em Phys.\ Rev.\ A 44} (1991), 682.

\bibitem{Spiller1993a}
{\sc Spiller, T.~P., Garraway, B.~M., and Percival, I.~C.}
\newblock {\em Phys.\ Lett.\ A 179} (1993), 63.

\bibitem{Spiller1994a}
{\sc Spiller, T.~P., and Ralph, J.~F.}
\newblock {\em Phys.\ Lett.\ A 194\/} (1994), 235.

\bibitem{Spiller1995a}
{\sc Spiller, T.~P., Ralph, J.~F., Clark, T.~D., Prance, R.~J., and Prance, H.}
\newblock The behaviour of individual ultra-small capacitance quantum devices.
\newblock submitted to Int.\ J. Mod.\ Phys.\ B.

\bibitem{Steinbach1994}
{\sc Steinbach, J., Garraway B.~M., and Knight, P.~L.}
\newblock {\em Phys.\ Rev.\ A 51} (1995), 3302.

\bibitem{Zheng1995}
{\sc Zheng, X.~P., and Savage, C.~M.}
\newblock {\em Phys.\ Rev.\ A 51} (1995), 792.

\end{thebibliography}
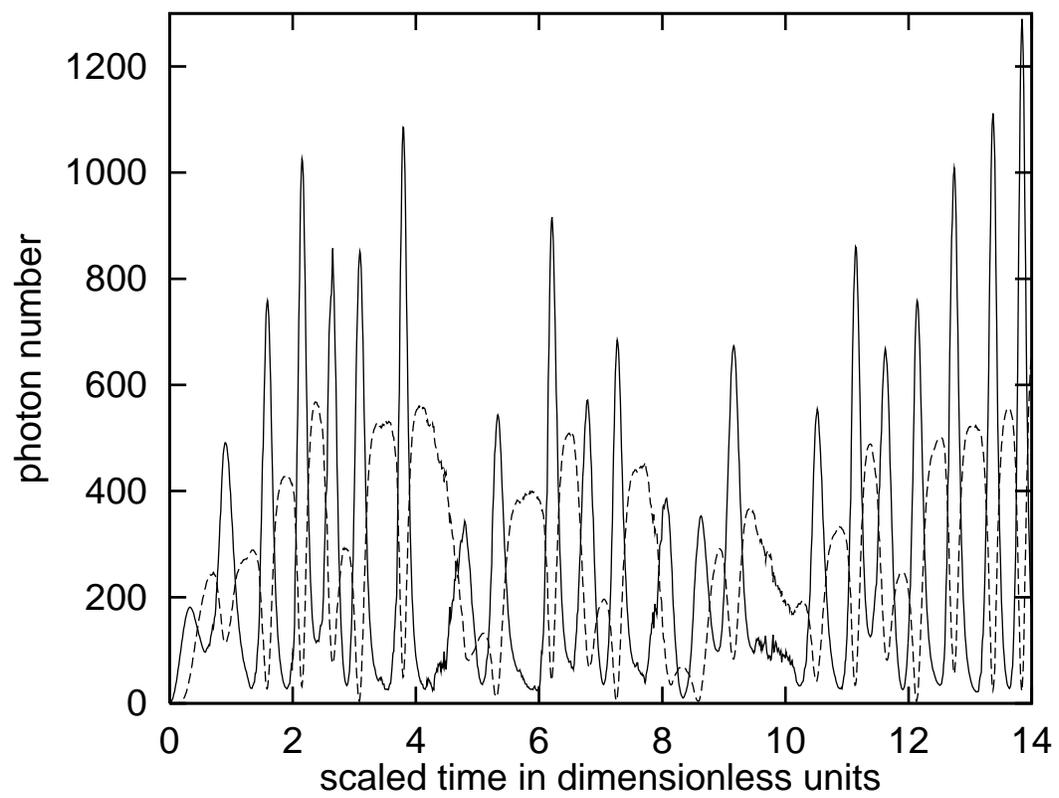
\end{document}